\documentclass[pra,reprint,amssymb]{revtex4-1}%

\usepackage{amsmath,amssymb}
\usepackage{revsymb4-1}
\usepackage[cmyk]{xcolor}
\usepackage[%
	colorlinks=true,%
	linkcolor=blue,%
	citecolor=blue,%
	urlcolor=blue
	]{hyperref}%
\usepackage{graphicx}%
\usepackage{here}
\usepackage{dcolumn}%
\usepackage{bm}%
\usepackage{revsymb}
\usepackage{braket}
\usepackage{layouts} %to get size.  \printinunitsof{in}\prntlen{\hsize}
\usepackage{color}

\def\equationautorefname~#1\null{\textrm{~(#1)\;}\null}
\def\figureautorefname~#1\null{Fig.~#1\null}

\begin{document}

\title{%
Nonlinear Mixing of Collective Modes in Harmonically Trapped Bose--Einstein Condensates
} 
\date{\today}

\author{Takahiro Mizoguchi}
\author{Shohei Watabe}
\author{Tetsuro Nikuni}

\affiliation{Tokyo University of Science, 1-3 Kagurazaka, Shinjuku-ku, Tokyo, 162-9601, Japan}

%\begin{spacing}{2} %%%% only for editing %%%%%

\begin{abstract}
We study nonlinear mixing effects among quadrupole modes and scissors modes in a harmonically trapped Bose--Einstein condensate.  Using a perturbative technique in conjunction with a variational approach with a Gaussian trial wave function for the Gross-Pitaevskii equation, we find that mode mixing selectively occurs.  Our perturbative approach is useful in gaining qualitative understanding of the recent experiment [Yamazaki {\it et al.}, J. Phys. Soc. Japan {\bf 84}, 44001 (2015)], exhibiting a beating phenomenon of the scissors mode as well as a modulation phenomenon of the low-lying quadrupole mode by the high-lying quadrupole mode frequency. Within the second-order treatment of the nonlinear mode coupling terms, our approach predicts all the spectral peaks obtained by the numerical simulation of the Gross-Pitaevskii equation. 
\end{abstract}

\maketitle
% \printinunitsof{pt}\prntlen{\hsize}
\section{Introduction}%%%

Collective modes, such as dipole, quadrupole and scissors modes, in harmonically trapped ultracold atomic gases have attracted much attention because they offer many-body physics as well as macroscopic quantum phenomena of Bose--Einstein condensates (BECs)~\cite{Jin1996,Mewes1996,Jin1997,Marago2000,Marago2001}.  In particular, the scissors mode in a BEC manifests superfluidity because of its irrotational nature~\cite{Marago2000,Marago2001,PhysRevLett.83.4452,Guery-Odelin1999a,Jackson2001}.  Frequencies of the collective modes observed in the experiment~\cite{Jin1996} can be successfully described by the linearized Gross-Pitaevskii (GP) equation for the condensate at $T=0$~\cite{Edwards1996,Stringari1996}.  Non-zero temperature effects~\cite{Guery-Odelin1999a,Nikuni2002,Jackson2001}  as well as the Beliaev process~\cite{Hodby2001} are also important for explaining experimental results~\cite{Jin1997,Marago2000,Marago2001,Hodby2001}, which are not explained by linearized equations at $T = 0$ alone. 

Even in the case at $T = 0$, however, we are faced with an important problem that are not explained by the linear analysis. Recently, various kinds of collective modes in a BEC---low-lying and high-lying quadrupole modes, as well as scissors modes---are simultaneously excited in a cigar-shaped trap~\cite{Yamazaki2015}. This experiment has shown that oscillations of a scissors mode exhibits beating with a longer period oscillation, whose frequency is that of the low-lying quadrupole mode. The low-lying quadrupole mode is, on the other hand, modulated by a shorter period oscillation with the high-lying quadrupole mode frequency. Since these phenomena cannot be described by the linearized GP equation, we are confronted with the following questions: (1) Why does the scissors mode exhibits beating with the low-lying quadrupole mode frequency? (2) Why is the low-lying quadrupole mode is modulated by the high-lying quadrupole mode frequency? (3) How is a collective mode possibly modulated by the other collective mode frequencies in a trapped Bose gas?  
These phenomena may be seen widely, since a similar phenomenon to Ref.~\cite{Yamazaki2015} has also been reported in a condensate response using a broadband probe with a wide range of frequency~\cite{Ruprecht1996}, where spectra at several sum and difference frequencies are slightly visible. 

Since these mode mixing effects are not included in the linear analysis, the phenomena we are interested in are beyond the linear analysis. In this paper, we study these nonlinear effects of collective modes by using a perturbative approach in conjunction with a variational calculation with a Gaussian trial wave function~\cite{Khawaja2001}. We show that the perturbative study successfully explains modulation phenomena of collective modes observed in the experiment of Ref.~\cite{Yamazaki2015}. The mode mixing effects shown in Ref.~\cite{Yamazaki2015} are concluded to be due to the nonlinear effects. Nonlinear mixing weight $G_{\rho\sigma\tau}'$ obtained in this paper clearly shows that mode mixing selectively occurs, and these weights answer all the three questions raised above. For the simultaneous excitation of quadrupole and scissors modes, by using the second-order perturbative treatment in the nonlinear mode coupling terms, we successfully predict all spectral peaks obtained from the numerical simulation of the GP equation.

\section{Variational analysis}

We derive nonlinear equations in order to understand nonlinear mixing of scissors and quadrupole modes.  The earlier study~\cite{Khawaja2001} considered nonlinear coupling between scissors modes with different symmetries.  In this section, we extend their variational analysis with a Gaussian trail wave function~\cite{Khawaja2001} to include mixing effects between both quadrupole and scissors modes in a trapped condensate. 

As in Ref.\cite{Khawaja2001}, we start with the Lagrangian for the condensate order parameter $\psi ({\bf r},t)$, given by  
\begin{eqnarray}
		L\left[\psi , \psi ^{*}\right]
	&= &
		\frac{i\hbar }{2}
		\int d{\bf r}
		\left(
			\psi ^{*}
			\frac{\partial \psi }{\partial t}
			-
			\psi \frac{\partial \psi ^{*}}{\partial t}
		\right)
		-E\left[\psi ,\psi ^{*}\right], 
	\label{defOfLagrangian}
\end{eqnarray}
where the energy functional is given by
\begin{eqnarray}
		E%\left[\psi ,\psi ^{*}\right]
	&= &
		\int d{\bf r}
		\left[
			\frac{\hbar ^{2}}{2m}
			\left|\nabla \psi \right|^{2}
			+V|\psi |^{2}
			+\frac{g}{2}\left|\psi \right|^{4}
		\right.
%\nonumber\\&&
		\left.
			-\mu \left|\psi \right|^{2}
		\right ] . 
	\label{energyFunctional}
\end{eqnarray} 
Here, $g=4\pi a\hbar ^{2}/m$ is a coupling constant, $a$ an $s$-wave scattering length, and $m$ an atomic mass. 
Its Euler-Lagrange equation is the GP equation 
\begin{align}
		\textrm{i}\hbar \frac{\partial \psi ({\bf r},t)}{\partial t}
	&=
		\left[
			-\frac{\hbar ^{2}\nabla ^{2}}{2m}
			+V({\bf r})
			+g\left|\psi ({\bf r},t)\right|^{2}
		\right]
		\psi ({\bf r},t). 
\label{GPE}
\end{align}
	The external trap potential is given by 
\begin{align}
		V({\bf r})
	&=
	\frac{m}{2}(
		\omega _x^{2}x^{2}
		+\omega _y^{2}y^{2}
		+\omega _z^{2}z^{2}
	)
\notag\\
	&=
	\frac{m}{2}
	\omega _\perp^{2}
	(
		\kappa ^{2}x^{2}
		+y^{2}/\kappa ^{2}
		+\lambda  ^{2}z^{2}
	),
\label{potential}
\end{align}
where $\omega _x$, $\omega _y$ and $\omega _z$ are trap frequencies of $x,y,z$ direction, respectively. The parameters $\kappa$, $\lambda$, and $\omega _\perp $ are given by $\omega _\perp = \sqrt{\omega _x\omega _y}$, $\kappa =\omega _x/\omega _y$, and $\lambda =\omega _z/\sqrt{\omega _x\omega _y}$. In this section, we describe equations in terms of  $\omega _x,\omega _y$ and $\omega _z$ to formulate general equations for arbitrary trap geometries. In the next section, we will use the parameters $\omega _\perp$ and $\lambda $ to study an axially symmetric trap case ($\kappa = 1$). 

We introduce the following trial function for the condensate order parameter~\cite{Khawaja2001}
\begin{align}
		\psi ({\bf r},t)
	=&
		A(t)
		\exp\left[
			-b_{x}(t)x^{2}
			-b_{y}(t)y^{2}
			-b_{z}(t)z^{2}
		\right.
	\notag\\&
		\left.
			-c_{xy}(t)xy
			-c_{yz}(t)yz
			-c_{zx}(t)zx
		\right],
	\label{defOfTrialFunc}
\end{align}
where $b_\zeta $, and $c_{\zeta \eta }$ ($\zeta, \eta =x,y,z$) are complex variational parameters. 
The parameter $A$ is a normalization factor to ensure the conservation of the total number of condensate atoms $N$, which is given by
\begin{align}
		A(t)
	=&
		\frac{2^{1/4}\sqrt{N}}{\pi ^{3/4}}
		[
			c_{xy,\textrm{r}}
			c_{yz,\textrm{r}}
			c_{zx,\textrm{r}}
			+4
			b_{x,\textrm{r}}
			b_{y,\textrm{r}}
			b_{z,\textrm{r}}
\notag\\&
			-
			(
				 b_{z,\textrm{r}}c_{xy,\textrm{r}}^{2}
				+b_{y,\textrm{r}}c_{zx,\textrm{r}}^{2}
				+b_{x,\textrm{r}}c_{yz,\textrm{r}}^{2}
			)
		]^{1/4},
	\label{Khawaja2001-3}
\end{align}
where $b_{\zeta ,\textrm{r}}$ and $c_{\zeta \eta ,\textrm{r}}$ ($b_{\zeta ,\textrm{i}}$ and $c_{\zeta \eta ,\textrm{i}}$) are real (imaginary) parts of $b_\zeta $ and $c_{\zeta \eta }$, respectively. 

We derive equations of motions for variational parameters for a BEC at $T=0$ in three steps. 
First, by inserting the trial function \autoref{defOfTrialFunc} into the Lagrangian \autoref{defOfLagrangian} and carrying out the spatial integration, we obtain
\begin{eqnarray}
		&&\frac{L[b,c]}{N}
	=
		\frac{1}{Q}
		\left(
			\sum_{\zeta =x,y,z}\alpha _\zeta  \dot b_{\zeta ,\textrm{i}}
			+\sum_{\{\zeta ,\eta ,\theta \}\in X} \beta _\zeta \dot c_{\eta \theta ,\textrm{i}}
		\right)
	\notag\\&&
		-\frac{1}{2Q}
			\sum_{\{\zeta ,\eta ,\theta \}\in X}
			\alpha _\zeta \left(
				4|b_\zeta |^{2}+|c_{\zeta \eta }|^{2}+|c_{\theta \zeta }|^{2}
			\right)
	\notag\\&&
		-\frac{1}{Q}
		\sum_{\{\zeta ,\eta ,\theta \}\in X}
		\sum_{\textrm{p}=\textrm{r},\textrm{i}}
		\beta _\zeta \left[
			2c_{\eta \theta ,\textrm{p}}\left(b_{\eta ,\textrm{p}}+b_{\theta ,\textrm{p}}\right)
			+c_{\zeta \eta ,\textrm{p}}c_{\theta \zeta ,\textrm{p}}
		\right]
	\notag\\&&
		-\frac{1}{2Q}\sum_{\zeta =x,y,z}\alpha _\zeta  \omega _\zeta  ^{2}
		-\frac{\gamma }{2\sqrt{\pi }}\sqrt{Q}, 
	\label{integratedLagrangian}
\end{eqnarray}
where $Q = 2\pi ^{3}A^{4}/N$, and $\gamma = Na/L_\textrm{ho}$. Here, $\alpha _\zeta \equiv 4b_{\eta ,\textrm{r}}b_{\theta ,\textrm{r}}-c^{2}_{\eta \theta ,\textrm{r}}$, and $\beta _\zeta \equiv c_{\theta \zeta ,\textrm{r}}c_{\zeta \eta ,\textrm{r}}-2b_{\zeta ,\textrm{r}}c_{\eta \theta ,\textrm{r}}$, where a set $\{ \zeta ,\eta ,\theta \}$ indicates Cartesian coordinates in cyclic order: $\{ \zeta ,\eta ,\theta \} \in \{\{x,y,z\}, \{y,z,x\}, \{z,x,y\}\}$. The dot represents a time derivative. All parameters have been scaled by the trap frequency $\bar\omega =(\omega _x\omega _y\omega _z)^{1/3} $ and the harmonic oscillator length $L_\textrm{ho}=(\hbar /m\bar\omega )$. The third term on the right hand side of (\ref{integratedLagrangian}) was missing in the earlier study~\cite{Khawaja2001}. 

Second, we look for static equilibrium values of variational parameters $b_\zeta ^\textrm{equiv}$ and $c_{\zeta \eta }^\textrm{equiv}$.
Since the phase of the static order parameter $\psi$ is uniform, we set $b_{\zeta ,\textrm{i}}^\textrm{equiv} = c_{\zeta \eta ,\textrm{i}}^\textrm{equiv} = 0$. Because of symmetry of the trap potential, we can set $c_{\zeta \eta ,\textrm{r}}^\textrm{equiv} = 0$. The equilibrium values of parameters $b_\zeta ^\textrm{equiv}$ can be obtained by minimizing the energy functional \autoref{energyFunctional}, which leads to
\begin{align}
		b_\zeta ^\textrm{equiv}
	=&
		\left(
			\frac{\sqrt{\pi }}{8\gamma }
		\right) ^{2/5}
		\omega _\zeta ^{2}
		\;\;\left(\zeta =x,y,z\right).
	\label{Khawaja2001_7}
\end{align} 
We then consider deviations of variational parameters $\delta b_\zeta (t) \equiv b_\zeta (t)-b_\zeta ^\textrm{equiv}$ and $\delta c_{\zeta \eta }(t) \equiv c_{\zeta \eta }(t) -  c_{\zeta \eta }^\textrm{equiv}$. 
It is useful to express these variational parameters in a vector form 
\begin{eqnarray}
	{\bf p}^{\textrm{T}}&\equiv&(\delta b_{x,\textrm{r}},\delta b_{y,\textrm{r}},\delta b_{z,\textrm{r}},\delta b_{x,\textrm{i}},\delta b_{y,\textrm{i}},\delta b_{z,\textrm{i}},
	\notag\\
	&&\delta c_{xy,\textrm{r}},\delta c_{yz,\textrm{r}},\delta c_{zx,\textrm{r}},\delta c_{xy,\textrm{i}},\delta c_{yz,\textrm{i}},\delta c_{zx,\textrm{i}}).
    \label{vectorp}
\end{eqnarray} 

Finally, we expand the Lagrangian in terms of ${\bf p}$, and derive the Euler-Lagrange equations to second order in ${\bf p}$, which can be written in the form
\begin{eqnarray}
		\textrm{i} M_{\mu \nu } \dot p_\nu 
		+ F_{\mu \nu} p_\nu 
		+ G_{\mu \nu \xi }p_\nu p_\xi 
		+ H_{\mu \nu \xi }p_\nu \dot p_\xi 
		=0.
	\label{eqsOfMotion1}
\end{eqnarray}
Here, $p_\nu $ is the $\nu $-th component of the vector ${\bf p}$, and we use the Einstein summation convention that implies taking the sum over the repeated indices. 
The coefficients $M_{\mu \nu }$ are pure imaginary, while the $F_{\mu \nu },G_{\mu \nu \xi }$ and $H_{\mu \nu \xi }$ are real.
The coefficients $M_{\mu \nu }$, $F_{\mu \nu }$, and $G_{\mu \nu \xi }$ are (a)symmetrised such as $M_{\mu \nu }=-M_{\nu \mu }$, $F_{\mu \nu }=F_{\nu \mu }$ as well as $G_{\mu \nu \xi }=G_{\mu \xi \nu }$. Note that $H_{\mu \nu \xi }$ does not have a symmetry. 
The matrix $M$ is given by
\begin{eqnarray}
		M
	&=&
		\textrm{i}\pi ^{1/5}
		\begin{pmatrix}
			0 &M^\textrm{q} &0 &0\\
			-M^\textrm{q} &0 &0 &0 \\
			0 &0 &0 &-M^\textrm{s} \\
			0 &0 &M^\textrm{s} &0 \\
		\end{pmatrix},
\end{eqnarray}
where
\begin{eqnarray}
		 M^\textrm{q}
	&=&
		\textrm{diag}(2\omega _y^{4}\omega _z^{4},\; 2\omega _z^{4}\omega _x^{4},\; 2\omega _x^{4}\omega _y^{4}),
\\
		 M^\textrm{s}
	&=&
		\textrm{diag}(\omega _z^{2},\; \omega _x^{2},\; \omega _y^{2}),
\end{eqnarray}
and $F$ is given by
\begin{eqnarray}
		F
	&=&
		\begin{pmatrix}
			F^\textrm{qr} &0 &0 &0\\
			0 &F^\textrm{qi} &0 &0 \\
			0 &0 &F^\textrm{sr} &0 \\
			0 &0 &0 &F^\textrm{si} \\
		\end{pmatrix},
\end{eqnarray}
where 
\begin{eqnarray}
		F^\textrm{qr}
	&&=
		2^{1/5}\gamma ^{2/5}
		\begin{pmatrix}
			3\omega _y^{4}\omega _z^{4} &\omega _z^{2} &\omega _y^{2} \\
			\omega _z^{2} &3\omega _z^{4}\omega _x^{4} &\omega _x^{2}	\\
			\omega _y^{2} &\omega _x^{2} &3\omega _x^{4}\omega _y^{4}
		\end{pmatrix}, 
\\
		F^\textrm{qi}
	&&=
		\frac{2^{9/5}}{\gamma ^{2/5}}
		\textrm{diag}(
			\omega _y^{2}\omega _z^{2},\;
			\omega _z^{2}\omega _x^{2},\;
			\omega _x^{2}\omega _y^{2}
			), 
\\
		F^\textrm{sr}
	&&=
		2^{1/5}\gamma ^{2/5}
		\textrm{diag}(
			\omega _z^{2},\;\omega _x^{2},\;\omega _y^{2}
		), 
\\
		F^\textrm{si}
	&&=
		\frac{\pi ^{2/5}}{2^{1/5}\gamma ^{2/5}}
		\textrm{diag}(
			\omega _z^{2}(\omega _x^{2}+\omega _y^{2}),\;
\notag\\&&\hphantom{=\frac{\pi ^{2/5}}{2^{1/5}\gamma ^{2/5}}}
			\omega _x^{2}(\omega _y^{2}+\omega _z^{2}),\;
			\omega _y^{2}(\omega _z^{2}+\omega _x^{2})
		).
\end{eqnarray}

\section{Perturbative Approach}

Within a linear approximation ignoring the nonlinear terms that includes $G$ and $H$ in \autoref{eqsOfMotion1}, we obtain the scissors and quadrupole mode frequencies, which are consistent with those in the hydrodynamic (strongly interacting or large-$N$) limit~\cite{Stringari1996,Khawaja2001}. Such a linear approximation is useful when we apply a single-frequency probe. However, if we excite collective modes simultaneously, nonlinear mode mixing may emerge and its effects are visible as in Refs.~\cite{Yamazaki2015,Ruprecht1996}. In order to study these effects beyond a linear analysis, we use the solution of a linear approximation as an unperturbed solution and regard the nonlinear terms as perturbation. To make this idea more concrete, we introduce a fictitious perturbation parameter $\epsilon$ such that 
\begin{align}
		\textrm{i} M_{\mu \nu } \dot p_\nu 
		+ F_{\mu \nu} p_\nu 
		+ \epsilon \left(
			  G_{\mu \nu \xi }p_\nu p_\xi 
			+ H_{\mu \nu \xi }p_\nu  \dot p_\xi 
		\right)
		=0.
		\label{eqsOfMotion}
\end{align}
Following a usual perturbation technique, we will set $\epsilon$ equal to unity at the end of calculation. We look for an approximate solution as a power series of $\epsilon$: ${\bf p}={\bf p}^{(0)}+\epsilon {\bf p}^{(1)}+\epsilon^2 {\bf p}^{(2)} + \cdots$. Substituting this power series into \autoref{eqsOfMotion} and comparing coefficients of each power of $\epsilon$, we obtain a series of simultaneous equations.

The unperturbed equation, the zeroth order of $\epsilon$, is given by 
\begin{align}
    \textrm{i} M_{\mu \nu } \dot p_\nu  ^{(0)}
		+ F_{\mu \nu } p_\nu ^{(0)} =0 . 
    \label{epsilonzero}
\end{align}
One can diagonalize this equation by a matrix $S$ such that $\dot q^{(0)}_\rho =  \textrm{i}  \Omega _{\rho } q^{(0)}_\rho $, where $ p^{(0)}_\mu  = S_{\mu \rho } q^{(0)}_\rho $ and $\Omega$ is a diagonal matrix given by $\Omega = S^{-1} M^{-1} F S$, whose elements are eigenvalues.  The eigenvalues are real numbers, each two of them having the same absolute value with opposite sign.  The solution $q^{(0)}_\rho $ can be given by $q^{(0)}_\rho  = c^{(0)}_\rho \exp{(\textrm{i} \Omega_{\rho } t)}$, where $c^{(0)}_\alpha  $ determines weights of eigenmodes in the initial condition. 

\begin{figure}[bt]
	\centering
	\includegraphics[width=1.0\hsize]{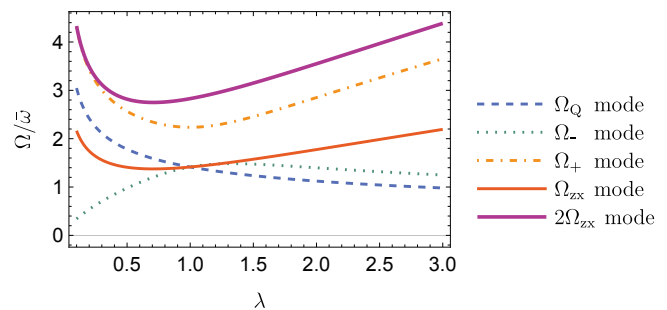}
	\caption{The $\lambda $-dependence of frequencies $\Omega _\textrm{Q}, \Omega _\pm, \Omega _+, \Omega _{zx}$ as well as $2\Omega _{zx}$ in an axially symmetric trap case ($\kappa =1$). The frequencies $\Omega _{yz}$ and $\Omega _{xy}$ are equal to the $\Omega _{zx}$ and $\Omega _\textrm{Q}$, respectively.
	}
	\label{fig:freqOfLinearSolution}
\end{figure}

In the following, we arrange eigenvalues such that 
\begin{widetext}
\begin{align} 
    { \Omega }=\textrm{diag} (-\Omega _\textrm{Q},\;\Omega _\textrm{Q},-\Omega _-,\Omega _-,-\Omega _+,\Omega _+, -\Omega _{xy},\Omega _{xy},-\Omega _{yz},\Omega _{yz},-\Omega _{zx},\Omega _{zx}). 
\end{align} 
\end{widetext}
In an axially symmetric geometry ($\kappa =1$), frequencies $\Omega_{\textrm{Q},\pm}$ correspond to the quadrupole mode frequencies, and frequencies $\Omega _{xy,yz,zx}$ correspond to the scissors mode frequencies, which are given by 
\begin{align} 
	\Omega _\textrm{Q}^{2} = & \Omega _{xy}^{2}=\frac{2}{\lambda ^{2/3}}
	,\nonumber \\
	\Omega _\pm^{2} = & \frac{4+3\lambda ^{2}\pm\sqrt{16-16\lambda ^{2}+9\lambda ^{4}}}{2\lambda ^{2/3}}
	, 
	\label{Omegasets} \\ 
	\Omega _{{zx}}^{2} = & \Omega _{yz}^{2}=\frac{1+\lambda ^{2}}{\lambda ^{2/3}}.
	\nonumber 
\end{align}  
The mode with $\Omega_{+(-)}$ is the high(low)-lying quadrupole-monopole excitation.  In an axially symmetric trap, the $\Omega_\textrm{Q}$ quadrupole mode and the $xy$ scissors mode degenerate, so do the $yz$ scissors mode and the $zx$ scissors mode. Figure~\ref{fig:freqOfLinearSolution} shows the $\lambda $-dependence of these frequencies.  These frequencies are consistent with those for a trapped condensate in the hydrodynamic limit~\cite{Stringari1996}.

\begin{figure*}[tb]
	\centering
	\includegraphics[width=450pt]{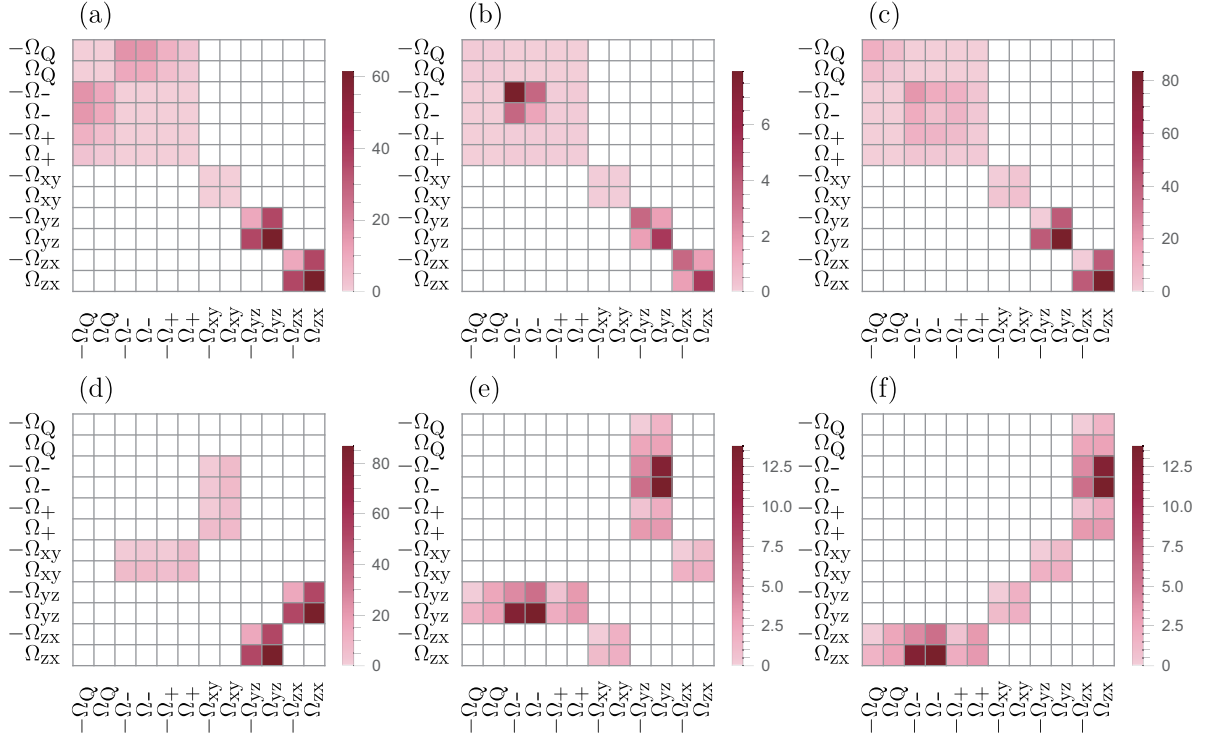}
    \caption{Absolute values of nonlinear mixing weight $G'_{\rho \sigma \tau }$ in the matrix form in a cigar-shaped trap case ($\lambda =0.14$). Panel (a) is the matrix $G'_{2,\sigma ,\tau }$($\Omega _\textrm{Q}$mode), (b) $G'_{4,\sigma ,\tau }$($\Omega _\textrm{-}$mode), (c) $G'_{6,\sigma ,\tau }$($\Omega _\textrm{+}$mode), (d) $G'_{8,\sigma ,\tau }$($\Omega _\textrm{xy}$mode), (e) $G'_{10,\sigma ,\tau }$($\Omega _\textrm{yz}$mode), and (f) $G'_{12,\sigma ,\tau }$($\Omega _\textrm{zx}$mode). The row and column of the matrix represent the $\sigma$ and $\tau $ indices, respectively. %
The $(\sigma,\tau )$ section of each matrix provides the modulation term with the frequency $\Omega _\sigma +\Omega _\tau $ for the $\rho$-th mode. For instance, the $(\sigma =1,\tau =12)$ section of each matrix provides the modulation term with $\Omega _1+\Omega _{12}=-\Omega _\textrm{Q}+\Omega _{\textrm{zx}}$. Blank (white color) elements are exactly zero for arbitrary $\lambda $, where the nonlinear mixing is absent. The absolute value of $G'_{\rho \sigma \tau }$ is equal to that of $G'_{\rho +1,\sigma +1,\tau +1}$ when $(\rho ,\sigma ,\tau )$ are odd.
	} 
		\label{fig:matrixGdash}
\end{figure*}

\begin{figure*}[tb]
	\centering
	\includegraphics[width=1.0\hsize]{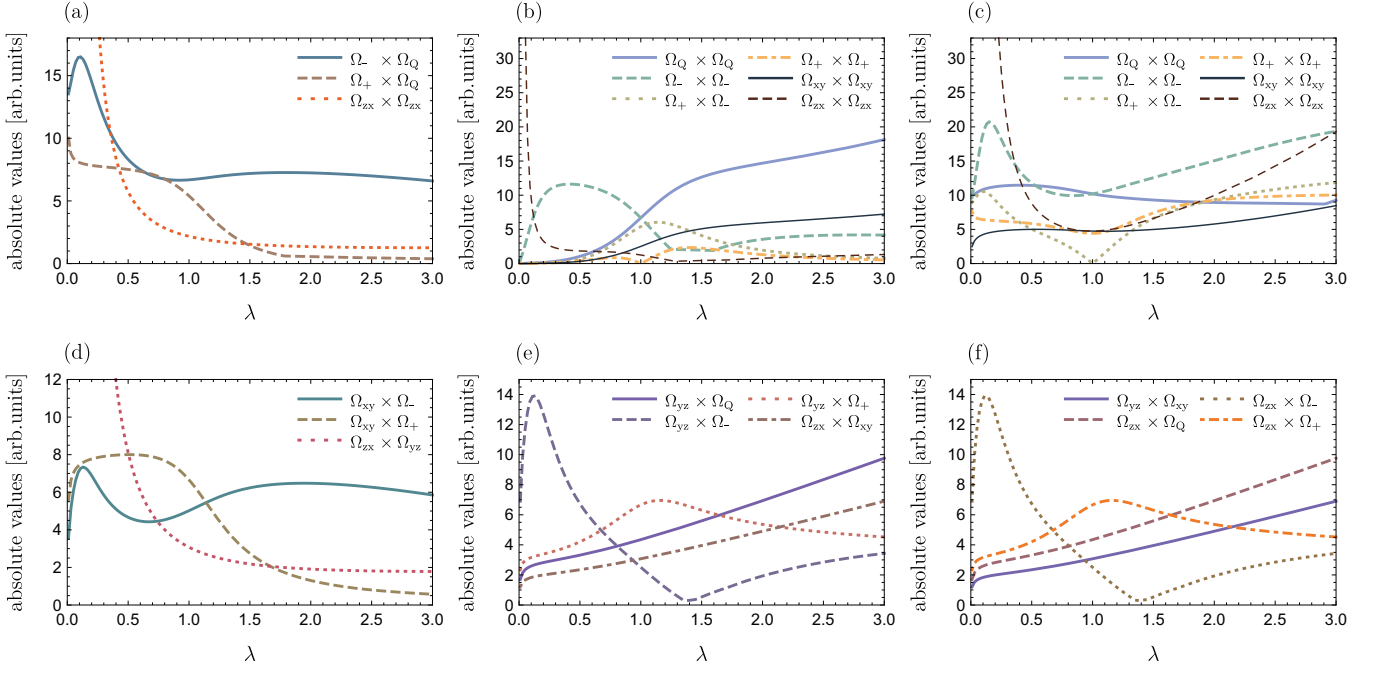}
	\caption{%
		The $\lambda $-dependence of mixing weight for (a) $\Omega _\textrm{Q}$ mode, (b) $\Omega _\textrm{-}$ mode, (c) $\Omega _\textrm{+}$ mode, (d) $\Omega _\textrm{xy}$ mode, (e) $\Omega _\textrm{yz}$ mode, (f) $\Omega _\textrm{zx}$ mode.
Each line is maximum absolute values among each four matrix elements in $G'$ for the $(\pm \Omega_j, \pm \Omega_k)$ sections as well as in the $(\pm \Omega_j, \mp \Omega_k)$ sections. Smaller value elements than those presented here are not shown. For instance, in (a), the line for $\Omega _-\times \Omega _\textrm{Q}$ is given by a maximum value among $|G'_{213}|,|G'_{214}|, |G'_{223}|, |G'_{224}|$ which correspond to $\Omega _\textrm{Q}\leftarrow -\Omega _{-}-\Omega _\textrm{Q}$, $\Omega _\textrm{Q}\leftarrow -\Omega _{-}+\Omega _\textrm{Q}$, $\Omega _\textrm{Q}\leftarrow +\Omega _{-}-\Omega _\textrm{Q}$, $\Omega _\textrm{Q}\leftarrow +\Omega _{-}+\Omega _\textrm{Q}$. In Figs. (a)-(c), lines for $\Omega _{yz}\times\Omega _{yz}$, which are not shown here, are the same as those for $\Omega _{zx}\times\Omega _{zx}$, because of an axially symmetric geometry case ($\kappa = 1$).
	   }
	\label{lambdaDependenciesOfGdashMaximum.eps}
\end{figure*}

Collecting terms of first order in $\epsilon$, we obtain 
\begin{align}
	\textrm{i} M_{\mu \nu } \dot p_\nu ^{(1)}
	+ F_{\mu \nu} p_\nu ^{(1)}
	+ f_{\mu }^{(0,0)}
	=0,
	\label{epsilonfrist}
\end{align} 
where $f_\mu ^{(i,j)}\equiv G_{\mu \nu \xi }p_\nu^{(i)} p_\xi ^{(j)}+ H_{\mu \nu \xi }p_\nu^{(i)}  \dot p_\xi ^{(j)}$. Using the transformation $S$, this equation reads as 
\begin{eqnarray}
		\dot q^{(1)}_\rho 
	=
		\textrm{i}\Omega _\rho 
		q^{(1)}_\rho 
		+
		\textrm{i}S^{-1}_{\rho \nu' }
		M^{-1}_{\nu' \mu }
		f^{(0,0)}_\mu
		,
    \label{epsilonfrist2}
\end{eqnarray}
where $ p^{(1)}_{\nu }\equiv S_{\nu \rho }{q}^{(1)}_{\rho }$.
The second term on the right hand side, $\textrm{i}S^{-1}_{\rho \nu' }M^{-1}_{\nu' \mu }f^{(0,0)}_\mu$, represents the nonlinear couplings between the collective modes, i.e. the eigenmodes in the linearized equation (\ref{epsilonzero}). In order to clearly understand nonlinear mixing effects of eigenmodes, it is more convenient to express the nonlinear term as $\textrm{i}S^{-1}_{\rho \nu' }M^{-1}_{\nu' \mu }f^{(0,0)}_\mu = G'_{\rho \sigma \tau }q^{(0)}_\sigma q^{(0)}_\tau $. From this expression, we will find that the $\rho  $-th mode is modulated by nonlinear mixing with the $\sigma $-th and $\tau  $-th modes through the term 
\begin{align}
	G'_{\rho \sigma \tau }
	c^{(0)}_{\sigma }
	c^{(0)}_{\tau }
	\exp[
		\textrm{i}
		(\Omega _\sigma  +\Omega _\tau  )
		t
	].
    \label{}
\end{align} 
This shows that the $\rho $-th mode is modulated by an oscillation with the frequency $\Omega _\sigma  +\Omega _\tau $, where we will present this effect as $\rho \leftarrow \sigma +\tau$ or $\Omega _\rho \leftarrow \Omega _\sigma +\Omega _\tau $.  Although the resultant nonlinear mixing may depend on an initial condition ${ c}^{(0)}_\rho (\rho =1,2,\cdots ,12)$, the weight $G'_{\rho \sigma \tau }$ is quite important in understanding the nonlinear mixing between eigenmodes in \autoref{epsilonzero}, because it determines the presence or absence of the nonlinear mixing. It is straightforward to find that the nonlinear mixing weight $G'$ is given by 
\begin{align}
		G'_{\rho \sigma\tau }
	&\equiv 
		\textrm{i}
		S^{-1}_{\rho \nu '}
		M^{-1}_{\nu '\mu }
\biggl [
				G_{\mu \nu \xi }
				S_{\nu \sigma }
				S_{\xi {\tau} } 
		\notag\\& 
			+
				\frac{\textrm{i}}{2}H_{\mu \nu \xi }
				\left(
				S_{\nu \sigma }S_{\xi \tau } \Omega _{\tau}
+
				S_{\nu \tau }S_{\xi \sigma } \Omega _{\sigma}
				\right)
		\biggr ]
,
\label{Mired}
\end{align}
where $G'$ is symmetric such as $G'_{\rho \sigma \tau }=G'_{\rho \tau \sigma }$. 

Figure \ref{fig:matrixGdash} shows absolute values of the nonlinear mixing weight $G'_{\rho }(\rho  =\textrm{even})$ in matrix form (column and row indices correspond to $\sigma$ and $\tau$, respectively) in a cigar-shaped trap case ($\lambda =0.14$, which corresponds to the experiment of Ref.\cite{Yamazaki2015}).  We do not show the cases where $\rho  $ is odd, because we have a relation $ G'_{\rho +1, \sigma +1, \tau +1 } = (G'_{\rho \sigma \tau })^*$ for $(\rho, \sigma, \tau )$ being odd, where this relation stems from a relation $q^{(1)}_{\rho  +1} = (q^{(1)}_{\rho  })^*$ for $\rho  $ being odd.  Some specific elements in matrices $G'_\rho  $ are absent, regardless of any value of the parameter $\lambda$, which indicates the absence of the nonlinear mixing of collective modes, as well as the selectiveness of the nonlinear mixing. 

Nonlinear mixing for the quadrupole modes is clearly distinct from that for the scissors modes.  Matrices $G'_2,G'_4$ and $G'_6$ for three quadrupole modes, which have essentially the same structure, have two characteristic features (Figs. \ref{fig:matrixGdash} (a), (b) and (c)).  First, all the quadrupole modes are coupled with each other, including itself.  This indicates that $\Omega_{\textrm{Q},\pm} \leftarrow \Omega_{\textrm{Q},\pm} \pm \Omega_{\textrm{Q},\pm} $. Second, the matrices $G'_2,G'_4$ and $G'_6$ involves a tridiagonal matrix, which leads to the nonlinear mixing $\Omega_{\textrm{Q},\pm} \leftarrow \Omega _{xy,yz,zx} + \Omega _{xy,yz,zx}$. The others such as $\Omega_{\textrm{Q},\pm} \leftarrow \Omega_{\textrm{Q},\pm} \pm \Omega _{xy,yz,zx} $ and $\Omega_{\textrm{Q},\pm} \leftarrow \Omega _{xy} \pm \Omega _{zx}$ do not occur. 

The matrices $G'_8, G'_{10}$ and $G'_{12}$ for the scissors modes have distinct two features (Figs. \ref{fig:matrixGdash} (d), (e) and (f)). First, the nonlinear mixing of all the quadrupole modes ($\Omega _{xy,yz,zx} \leftarrow \Omega_{\textrm{Q},\pm} \pm \Omega_{\textrm{Q},\pm} $) are absent. Second, the nonlinear mixing between a scissors mode and itself ($\sigma \leftarrow \sigma + \tau $ for $\sigma, \tau = 7,8, \cdots, 12$) are also absent . These two features are in contrast with those for $G'_2,G'_4$ and $G'_6$. 

On the other hand, a common feature can be found with respect to a beating phenomenon. From all the matrices $G'_2,G'_4,G'_6,G'_8,G'_{10}$ and $G'_{12}$, we find that the beating effect is caused by the quadrupole modes. Indeed, for the $\sigma$-th mode, we have a nonlinear mixing term with $\sigma \leftarrow \sigma + \tau$ for $\Omega_\tau = \Omega _\textrm{Q}$ and $\Omega _\pm$, where an exception is for the scissors mode in the $x$-$y$ plane, the coupling with the quadrupole mode ($\Omega_\tau=\Omega _\textrm{Q}$) being absent.  

The nonlinear mixing weight $G'$ is helpful in understanding the recent experiment \cite{Yamazaki2015}.  In this experiment, the $zx$ scissors mode with the frequency $\Omega_{zx}$ shows a beating phenomenon, where its oscillation amplitude is modulated by the longer-period oscillation with the low-lying quadrupole mode frequency $\Omega_-$. For the $zx$ scissors mode, the matrix elements of $G'_{12,\sigma ,\tau }$ for the $(\Omega_{zx}, \pm \Omega_-)$ sector are quite large compared with the other matrix elements (\autoref{fig:matrixGdash} (f)). This indicates that the nonlinear mixing between the $zx$ scissors mode ($\Omega_{zx}$) and the low-lying quadrupole mode ($\pm \Omega_{-}$) is strong, and the beating emerges from the term $\exp[\textrm{i} (\Omega_{{zx}} \pm \Omega_-) t]$. This fact clearly explains the experimental results \cite{Yamazaki2015}. 

Our perturbative approach also answers the question why the oscillation of the low-lying quadrupole mode was modulated by the short period oscillation with the high-lying quadrupole mode frequency ($\Omega_+$). For the low-lying quadrupole mode $(\Omega_{-})$, we can find relatively large matrix elements in $G'_{4,\sigma ,\tau }$ for the $(\Omega_{yz(zx)},\Omega_{yz(zx)})$ sections (\autoref{fig:matrixGdash} (b)). This indicates that the low-lying quadrupole mode is modulated by the oscillation $\exp{[\textrm{i}2\Omega_{yz(zx)}t]}$. Since the high-lying quadrupole mode frequency satisfies $\Omega_+ \simeq 2 \Omega_{yz(zx)}$ in a cigar-shaped trap case ($\lambda \ll 1$) (see \autoref{fig:freqOfLinearSolution}), which is the case in the experiment~\cite{Yamazaki2015}, the modulation of the low-lying quadrupole mode may be due to the nonlinear mixing between the $yz(zx)$ scissors modes, not due to the mixing between the low-lying and high-lying quadrupole modes. If the modulation were caused by the nonlinear mixing between these low-lying and high-lying quadrupole modes, a beating phenomenon could emerge like the scissors mode case. However, it was not the case in the experiment~\cite{Yamazaki2015}, where beating phenomena were not observed in the low-lying quadrupole mode. This result can be supported by the matrix element in the $(\Omega_-,\Omega_+)$ section quite smaller than that in the $(\Omega_{yz(zx)},\Omega_{yz(zx)})$ section (\autoref{fig:matrixGdash} (b)). 

Figure \ref{lambdaDependenciesOfGdashMaximum.eps} shows the $\lambda $-dependence of the dominant nonlinear mixing weight in $G'$. We plot the maximum absolute value among four matrix elements in the $(\pm \Omega_\sigma, \pm \Omega_\tau)$ sections as well as in the $(\pm \Omega_\sigma, \mp \Omega_\tau)$ sections. In the cigar shape trap case ($\lambda \ll 1$), the dominant nonlinear mixing for the quadrupole mode $\Omega_{\textrm{Q},\pm}$ is the mixing between the $zx$ scissors modes and itself, leading to the frequency $2\Omega_{zx}$ (Figs. \ref{lambdaDependenciesOfGdashMaximum.eps} (a), (b) and (c)). The dominant nonlinear mixing for the $yz(zx)$ scissors mode is the mixing between the $yz(zx)$ scissors mode and the low-lying quadrupole mode, leading to the frequency $\Omega _{yz(zx)}\pm\Omega _-$ (Figs. \ref{lambdaDependenciesOfGdashMaximum.eps} (e), and (f)). This is consistent with the experimental result~\cite{Yamazaki2015}. On the other hand, in a pancake-shaped trap case ($\lambda \gg 1$), the dominant nonlinear mixing for the $\Omega_\textrm{Q}$ quadrupole mode is the mixing between its own mode and the low-lying quadrupole mode. 

We can also discuss frequencies in the second-order perturbation by using the nonlinear mixing terms $G'_{\rho \sigma \tau }$. In the second order of $\epsilon$, we have 
\begin{align}
	\textrm{i} M_{\mu \nu } \dot p_\nu ^{(2)}
	+ F_{\mu \nu} p_\nu ^{(2)}
	+ f_\mu ^{(0,1)}
	+ f_\mu ^{(1,0)}
	=0.
\end{align} 
The inhomogeneous terms, $\textrm{i} S_{\rho \nu '}^{-1}M_{\nu '\mu }^{-1}f_\mu ^{(i,j)}$ for $(i,j) = (1,0)$ and $(0,1)$, have the same form as that in the first order case (\ref{epsilonfrist2}).  In the case $(i,j) = (1,0)$, the nonlinear mixing is given by ${G'}_{\rho \sigma \tau }^{(1,0)}q_\sigma ^{(1)}q_\tau ^{(0)}$, where $ {G'}_{\rho \sigma \tau }^{(1,0)}$ is the same as $G'_{\rho \sigma \tau }$ in \autoref{Mired}. On the other hand, in the case $(i,j) = (0,1)$, the nonlinear mixing is given by ${G'}_{\rho \sigma \tau }^{(0,1)}q_\sigma ^{(0)}q_\tau ^{(1)}$, where ${G'}_{\rho \sigma \tau }^{(0,1)}$ is defined by replacing the frequencies $\Omega _\sigma$ and  $\Omega _\tau $ in \autoref{Mired} with frequencies of $q_\sigma^{(1)}$ and $q_\tau^{(1)} $, respectively. Although values of ${G'}^{(0,1)}_{\rho \sigma \tau }$ itself may change from those given in \autoref{Mired}, the position of matrix elements being zeros, where nonlinear mixing is absent, does not change. This aspect is very useful in predicting the modulation frequency in the second order perturbation.  For instance, in \autoref{fig:matrixGdash} (a) for the matrix $G'_{2\sigma \tau }$, the index of the $\Omega _\textrm{Q}$ row may be replaced with $\Omega _\textrm{Q}+\Omega _-$, since we have the modulation $\Omega_\textrm{Q} \leftarrow \Omega_\textrm{Q} + \Omega_-$ in the first-order perturbation. In the second order perturbation, the modulation of the oscillation for the $\Omega_\textrm{Q}$ quadrupole mode then have the frequency with $2\Omega _\textrm{Q}+\Omega _-$ and $\Omega _\textrm{Q}+2\Omega _-$, i.e., the modulations $\Omega _\textrm{Q} \leftarrow (\Omega _\textrm{Q}+\Omega _- ) + \Omega _\textrm{Q} $ as well as $\Omega _\textrm{Q} \leftarrow (\Omega _\textrm{Q}+\Omega _- ) + \Omega _- $ occur.  On the other hand, in the same order perturbation for the $\Omega_\textrm{Q}$ quadrupole mode, the modulation with the frequency $2\Omega_-+\Omega _{zx}$, i.e., the modulation  $\Omega _\textrm{Q} \leftarrow (\Omega _- + \Omega _{zx} ) + \Omega _- $, may be absent, since in the first order perturbation, the modulation  $\Omega _\textrm{Q} \leftarrow \Omega _- + \Omega _{zx} $ does not exist (\autoref{fig:matrixGdash} (a)).

\section{Comparison with spectral analysis} 

\begin{figure}[p]
	\centering
	\includegraphics[width=246pt]{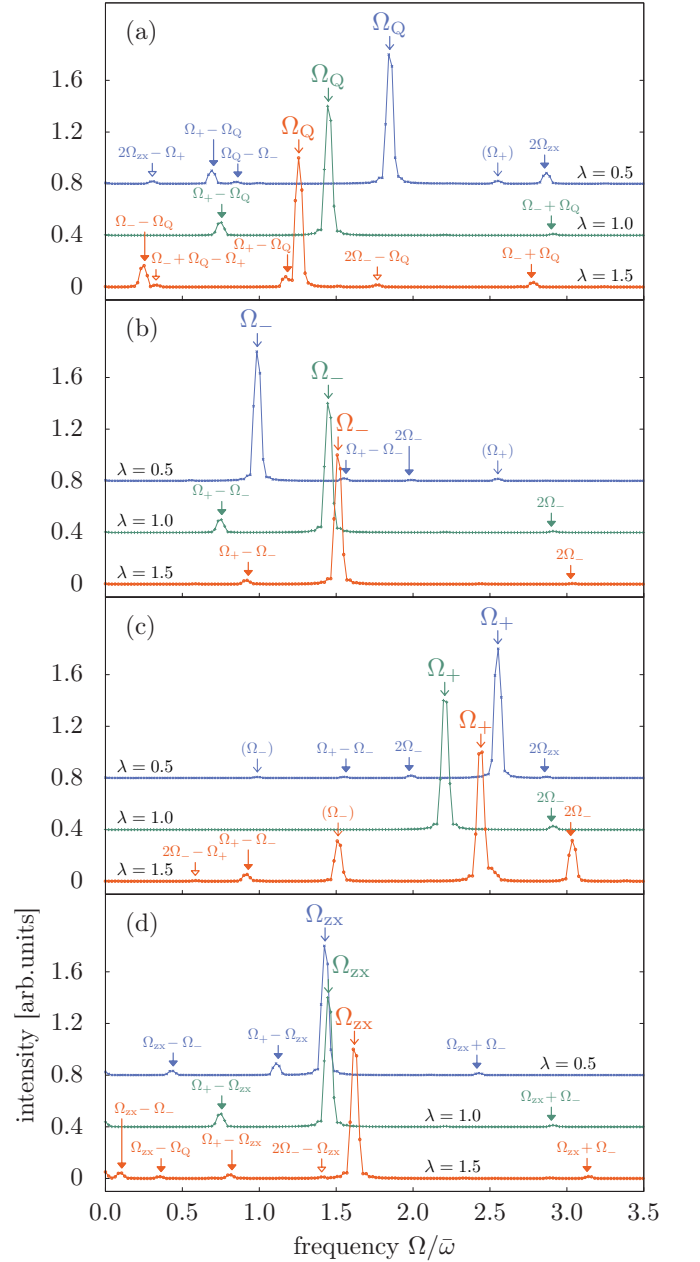}
	\caption{%
        Spectral intensities of modes for (a) $\Omega _\textrm{Q}$ mode, (b) $\Omega _-$ mode, (c) $\Omega _+$ mode, and (d) $\Omega _{zx}$ mode, obtained from numerical simulation of GP equation in axially symmetric trap cases ($\lambda =0.5,1.0,1.5$). Arrows with largest labels in each panel indicate frequencies $\Omega_{\textrm{Q},\pm, zx}$ that are determined by the maximum peak position. Filled-head (open-head) arrows together with a linear combination of the $\Omega_i$ label represent frequencies predicted by the first-order (second-order) perturbation approximation.  Spectrum of a quadrupole mode involves other quadrupole mode peak indicated by an arrow with a parenthetical label $(\Omega_i)$ (Figs. (a)-(c)), because each collective mode is extracted from the linear approximations such as (\ref{eq28})-(\ref{eq29}) by using moments $\braket{x^2}$,$\braket{y^2}$ and $\braket{z^2}$ obtained from the nonlinear numerical simulation. 
    } 
	\label{fig:multiplotOfSimulations.eps}
\end{figure}

We compare the analytical results in the previous section with the numerical simulation of the GP equation.
In this simulation, we simultaneously excite the quadrupole modes and the scissors mode using a perturbative external potential $\delta V({\bf r},t)=\theta (-t)(0.2x^{2}+0.25y^{2}+0.1zx)$ in the dimensionless form, where $\theta (t)$ is the Heaviside step function. This type of perturbative potential was used in the experiment of Ref.~\cite{Yamazaki2015} to excite three kinds of the quadrupole modes with $\Omega_{\textrm{Q},\pm}$ and the $zx$ scissors mode with $\Omega_{zx}$. 

To extract the spectra of each collective mode, we employ the following three steps. First, we calculate moments~\cite{Guery-Odelin1999a,PhysRevLett.83.4452,griffin2009bose} from the condensate wavefunction $\Psi({\bf r}, t)$, where the GP equation was numerically solved by applying the method in Ref.~\cite{Jackson2002}. These moments are defined as $\braket{\chi } \equiv \int d{\bf r} \chi \left|\Psi({\bf r},t) \right|^{2}$, where $\chi=x^{2}$, $y^2$, $z^2$, $xy$, $yz$, and $zx$. From these quantities, we evaluate deviations from their time-averaged values.  Second, we relate these moments to the vector ${\bf p}$ in (\ref{vectorp}) within the linear approximation. One finds that the explicit relations are given by 
\begin{align}
 \braket{x^{2}} \simeq &
 -\left(\frac{2}{\pi }\right)^{2/5}
		\frac{\gamma  ^{4/5}}{\omega _x^{4}}
		\delta b_{x,\textrm{r}}, 
        \label{eq28}
\\
\braket{xv_x}  \simeq &
		\frac{2}{b_x^{(0)}}\delta b_{x,\textrm{i}},
                \label{eq29}
\\
\braket{xy}  \simeq &
		-\frac{\gamma ^{4/5}}{2^{3/5}\pi ^{2/5}\omega _x^{2}\omega _y^{2}}
		\delta c_{xy,\textrm{r}}                \label{eq30},
\\
\braket{xv_y+yv_x}  \simeq &
		\left(\frac{1}{b_x^{(0)}}+\frac{1}{b_y^{(0)}}\right)\delta c_{xy,\textrm{i}}. 
                        \label{eq31}
\end{align} 
The analogous relations for the other moments are easily obtained. These relations enable us to give the vector ${\bf p}$ as a function of the moment $\braket{\chi}$, instead of the variational parameters $\delta b_\nu$ and $\delta c_{\mu\nu}$.  Finally, we apply the Fourier transformation to the vector ${\bf q} (\braket{\chi}) \equiv S^{-1} {\bf p} (\braket{\chi})$. 

Figure~\ref{fig:multiplotOfSimulations.eps} shows spectral intensities obtained from the numerical simulation in the GP equation. 
All peaks are excellently explained by our perturbative approach within the second order analysis (\autoref{fig:multiplotOfSimulations.eps}). 
One may find a relatively weak peak at the high-lying quadrupole mode frequency $\Omega_+$ in the $\Omega_\textrm{Q}$ quadrupole mode (\autoref{fig:multiplotOfSimulations.eps} (a)). This is because in order to extract each collective mode, we apply the linear approximations such as (\ref{eq28})-(\ref{eq29}) to $\langle \chi\rangle$ that includes all the nonlinear effects. This is the case in Figs. \ref{fig:multiplotOfSimulations.eps} (b) and (c). 

In the comparison of our perturbation approach with the numerical simulation, we use renormalized frequencies $\Omega_i$ that are determined from the principal spectral peaks in \autoref{fig:multiplotOfSimulations.eps}. This is because the frequencies in (\ref{Omegasets}) are rather consistent with the hydrodynamic (strongly interacting or large-$N$) limit, where the kinetic pressure energy is negligible~\cite{Stringari1996}, and it is not the case in this numerical simulation. In Appendix \ref{AppendixA}, we show results where the frequencies given in \autoref{Omegasets} are used. In the intermediate regime (not in the hydrodynamic regime), it is difficult to apply analytic calculations, but collective mode frequencies obtained in the experiment~\cite{Jin1996} are well described by the numerical calculation of the linearized equation~\cite{Edwards1996}. Since eigenmodes numerically obtained from the linearized equation in the intermediate regime \cite{Edwards1996} are smoothly connected to those in the hydrodynamic limit~\cite{Stringari1996}, our perturbative approach is applicable beyond the hydrodynamic limit if we take renormalized frequencies. This is the reason why our perturbative approach well predicts all the spectral peaks obtained in the numerical simulation not in the hydrodynamic limit. The merit of our present approach using the nonlinear mixing weight is that it gives clear physical understanding of nonlinear mixing effects between specific collective modes, such as quadrupole modes as  well as scissors modes.

\section{Conclusion} 
To understand nonlinear mixing effects among collective modes in a harmonically trapped Bose--Einstein condensates, we studied quadrupole modes and scissors modes by using a variational calculation with a Gaussian trial wave function together with a perturbation approach.  We derived equations of motions for variational parameters to second-order in fluctuations, and applied a perturbative approach to reveal structure of nonlinear couplings between unperturbed collective modes. We estimated the nonlinear mixing weight  and found that mode mixing selectively occurs. 

Although it is not always guaranteed to apply perturbative techniques to nonlinear equations, such as the Gross-Pitaevskii equation, our approach clearly explained the recent experiment~\cite{Yamazaki2015}, where the scissors mode exhibits a beating phenomenon with a longer period oscillation that corresponds to the low-lying quadrupole mode frequency, and an oscillation of the low-lying quadrupole mode is modulated by a shorter period oscillation with the high-lying quadrupole mode frequency. All the spectral peaks numerically obtained from the Gross-Pitaevskii equation are also excellently explained with the application of the second order perturbation analysis despite the case where the hydrodynamic limit approach is not applied. Nonlinear mixing weight obtained in our perturbation study will be helpful in understanding nonlinear mixing effects between collective modes in a trapped Bose--Einstein condensates.

\begin{acknowledgments} 
    The authors would like to thank M. Yamazaki for useful discussion. The second author is partially supported by JSPS KAKENHI Grant Number JP16K17774. 
\end{acknowledgments}

\appendix

\section{}\label{AppendixA}
\begin{figure}[h]
	\centering
	\includegraphics[width=246pt]{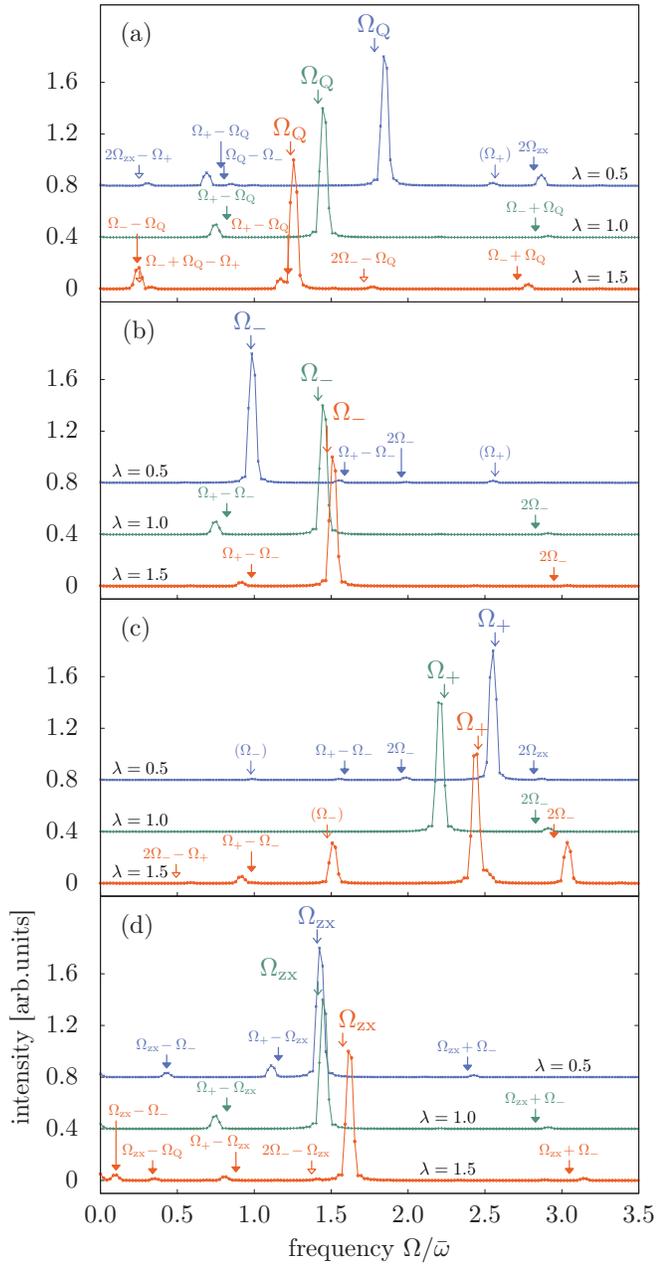}
	\caption{%
        Spectral intensities of modes for (a) $\Omega _\textrm{Q}$ mode, (b) $\Omega _-$ mode, (c) $\Omega _+$ mode, and (d) $\Omega _{zx}$ mode, together with arrows pointing the unperturbed frequencies \autoref{Omegasets}. Only the difference between \autoref{fig:multiplotOfSimulations.eps} and \autoref{fig:multiplots2_analyticFreq.eps} is whether to use the renormalized frequencies or the unperturbed frequencies \autoref{Omegasets} for arrows. 
    } 
	\label{fig:multiplots2_analyticFreq.eps}
\end{figure}
We show results of spectral analysis with the frequencies in \autoref{Omegasets} (\autoref{fig:multiplots2_analyticFreq.eps}). The situation of the numerical simulation is the same as that in \autoref{fig:multiplotOfSimulations.eps}. However, arrows are pointed based on the frequencies \autoref{Omegasets}, instead of the renormalized frequencies $\Omega_i$ determined from frequencies giving maximum peaks of intensities. In contrast to the case in \autoref{fig:multiplotOfSimulations.eps}, the arrows cannot well predict the peak position of the spectra. This is because the frequencies in \autoref{Omegasets} obtained in the variational approach correspond to those in the hydrodynamic (strongly interacting or large-$N$) limit~\cite{Khawaja2001}, which is no the case in the numerical simulation. 
 
\bibliographystyle{apsrev4-1}
\bibliography{library.bib}

\end{document}